# SENTIMENT-AWARE RECOMMENDATION SYSTEMS IN E-COMMERCE: A REVIEW FROM A NATURAL LANGUAGE PROCESSING PERSPECTIVE

AUTHOR

Yogesh Gajula

*University of Dayton*
*Dayton OH USA*
yogesh1297@gmail.com

## ABSTRACT

E-COMMERCE PLATFORMS generate vast volumes of user feedback, such as star ratings, written reviews, and comments. However, most recommendation engines rely primarily on numerical scores, often overlooking the nuanced opinions embedded in free text. This paper comprehensively reviews sentiment-aware recommendation systems from a natural language processing perspective, covering advancements from 2023 to early 2025. It highlights the benefits of integrating sentiment analysis into e-commerce recommenders to enhance prediction accuracy and explainability through detailed opinion extraction. Our survey categorizes recent work into four main approaches: deep-learning classifiers that combine sentiment embeddings with user–item interactions, transformer-based methods for nuanced feature extraction, graph neural networks that propagate sentiment signals, and conversational recommenders that adapt in real-time to user feedback. We summarize model architectures and demonstrate how sentiment flows through recommendation pipelines, impacting dialogue-based suggestions. Key challenges include handling noisy or sarcastic text, dynamic user preferences, and bias mitigation. Finally, we outline research gaps and provide a roadmap for developing smarter, fairer, and more user-centric recommendation tools.

PLAIN-LANGUAGE SUMMARY: Online shopping sites usually use star ratings to suggest products, but they miss out on useful information from written reviews. This paper looks at ways to mix machine reading and sentiment analysis to improve product suggestions. It discusses recent studies (2023–2025) using advanced text models to understand customer likes, talks about common review data sources, and explains testing methods like how accurate the suggestions are and how happy users are. Challenges include casual language, not enough data in niche markets, and supporting multiple languages. Promising methods use large pre-trained text models, but there's a need to reduce bias and speed up responses. The paper offers advice for creating better, fairer recommendation tools.

This paper is 6,618 words (excluding Abstract, References, Tables, and Figure Legends)

## 1 INTRODUCTION

Online recommendation systems have become a cornerstone of e-commerce, guiding users to products or content that match their preferences. Traditional recommender algorithms largely rely on numeric ratings or past behavior, but they often fail to capture the nuanced **context and intent** behind a user's choices. In modern e-commerce, **user-generated text** – especially product reviews and comments – contains rich **sentiment information** about likes, dislikes, and opinions. For example, a user review stating *"The phone camera is great, but the battery life is disappointing"* reveals specific positive and negative aspects that a simple star rating cannot convey. **Sentiment-aware recommendation systems** aim to incorporate these textual sentiments into the recommendation process, enabling more accurate and personalized suggestions. The intuition is that understanding *why* a user liked or disliked an item (through sentiment analysis of reviews) can help predict what other items they will enjoy. Moreover, leveraging sentiments can improve the *explainability* of recommendations, e.g. by highlighting relevant user opinions as part of the recommendation rationale [13].

**Natural Language Processing (NLP)**, and sentiment analysis in particular, provides the tools to extract and quantify opinions from text. In the past decade, sentiment analysis techniques have evolved from simple lexicon-based methods to sophisticated deep neural networks and transformer-based language models. This evolution has opened new opportunities to blend NLP with recommender systems. E-commerce giants like Amazon and Alibaba have access to millions of product reviews – for instance, the Amazon Reviews 2023 dataset contains over **570 million** reviews spanning 33 product categories [7]. Harnessing such data can significantly augment user profiles and item descriptions with sentiment signals. Recent research shows that incorporating review text leads to more personalized, accurate, and even **explainable** recommendations compared to using ratings alone [13][5].

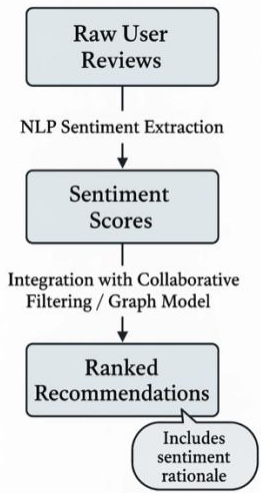

*Figure 1*
*High-level overview of an end-to-end sentiment-aware recommendation pipeline.*

**Motivation:** With the rapid growth of deep learning and language models, the period 2023–2025 has seen a surge of interest in sentiment-aware recommenders. Notably, there is a trend toward using **pre-trained transformer models** (like BERT, RoBERTa, GPT) to better understand review text and user emotions [10]. Researchers combine knowledge graphs with sentiment analysis to enhance context and resolve ambiguities in user opinions. Conversational recommender systems use natural language to respond to user emotions, improving recommendations by addressing dissatisfaction and ignoring negative feedback in reviews. [8].

*To ground the motivation above, Figure 1 shows how raw user reviews are first processed by an NLP sentiment-extraction module, then merged with a collaborative-filtering or graph engine to yield ranked recommendations enriched with sentiment rationale.*

**Scope of this review:** This paper provides a thorough overview of sentiment-aware recommendation systems in e-commerce, focusing on academic work from 2023 to early 2025. We organize the literature by methodological trends: (1) deep learning models that integrate textual sentiment features, (2) transformer-based and large language model approaches, (3) graph-based methods that incorporate sentiment knowledge, and (4) conversational systems that adapt to user sentiment in real-time dialogue. We summarize key models and their contributions in **Table 1** (Section 2) and highlight commonly used **datasets** in **Table 2** (Section 4). In each area, representative techniques are discussed, and **Figure 1** and **Figure 2** illustrate how sentiment information is utilized within a recommendation pipeline and a conversational scenario, respectively. We also discuss the **open challenges** in this field – such as handling noisy or sarcastic language, keeping models updated with evolving user sentiment trends, and ensuring fairness and robustness. Finally, we outline future research directions that may support continued progress and inspire further innovation in the development of sentiment-aware recommender systems.

Overall, our goal is to unite NLP and recommender systems by exploring sentiment analysis's role in improving e-commerce recommendations. This paper reveals state-of-the-art techniques, gaps, and future steps for user-feeling-aware engines, enhancing human-centric online shopping experiences.

# 2 LITERATURE SEARCH METHODOLOGY: APPROACHES TO SENTIMENT-AWARE RECOMMENDATIONS

A systematic search of IEEE Xplore, ACM Digital Library, Scopus, and arXiv for English-language publications on sentiment-aware recommendation systems between January 2023 and March 2025, using search terms such as "sentiment-aware recommender," "e-commerce review," and "natural language processing." After removing duplicates, we screened titles and abstracts against inclusion criteria—peer-reviewed or preprint articles focused on applying sentiment analysis to e-commerce recommenders—and reviewed full texts of eligible studies to assemble the final set for synthesis.

**Major Trends (2023–2025):** Recent literature reflects a convergence of advances in NLP and recommender systems. One clear trend is the shift from using manually crafted features or simple models to deploying **deep neural networks** that can learn complex representations of text and user preferences. Early sentiment-aware recommenders used techniques like topic modeling or sentiment lexicons to incorporate review information, but current systems more often leverage end-to-end deep learning [4]. In particular, **transformer-based language models** have gained prominence due to their superior ability to understand context and extract nuanced sentiment signals from text [4]. Another trend is the incorporation of **auxiliary information** through graph-based frameworks – for instance, building knowledge graphs of users, items, and opinion entities (aspects, sentiment words) to better capture relationships. There is also growing interest in **interactive and conversational systems** that can dynamically adjust recommendations based on a user's utterances and emotional cues during a dialogue. Table 1 summarizes several representative models and methods that exemplify these trends, which we detail in the following subsections.

| Model / Approach | Year | Key Idea & Methodology | Notable Findings |
|---|---|---|---|
| **RAKCR** [1] | 2024 | Review-aware Knowledge Graph Convolutional Network – integrates **sentiment weights** into user–item–entity graph. Utilizes GCN with sentiment-specific relational edges. | Captures fine-grained user/item features from sentiment-rich reviews, improving personalizations. Outperforms vanilla GCN on rating prediction in experiments. |
| **BERT for Rec Sys & Sentiment Analysis** [3] | 2024 | Hybrid recommender combining **BERT-based sentiment analysis** with collaborative filtering. Uses BERT to encode review text from Yelp (restaurants & hotels) into sentiment features, which are fed into a rating prediction model. | Incorporating review sentiment yielded higher precision/recall than using ratings alone. Demonstrated improved recommendations in two domains by leveraging opinion context. |
| **SENGR** [18] | 2022 | Sentiment-Enhanced Neural Graph Recommender – employs a **graph neural network (GNN)** to propagate both user-item interactions and sentiment signals (from review text). | One of the first GNN-based recommenders to include sentiment. Reported better performance on social recommendation tasks by modeling sentiment relationships |
| **SECR** [8] | 2025 | Sentimentally Enhanced Conversational Recommender – a **conversational RS** that analyzes user's emotional tone towards entities in dialogue. Builds a movie knowledge graph (MAKG) and filters entities by sentiment scores (using a sentiment lexicon and prompt-based analysis) to inform recommendations. | By detecting and prioritizing entities that align with the user's sentiment, the system avoids recommending items the user disliked in conversation. Achieved higher user satisfaction in dialog-based recommendation (movies domain). |
| **Chat-Rec** [20] | 2023 | LLM-augmented Recommender – integrates a **large language model (ChatGPT)** into the recommendation loop. The LLM performs sentiment analysis on user reviews and generates interactive recommendation explanations. | Improved cold-start recommendations by using GPT to infer user intent from minimal data Also produced more natural, explainable interactions, though with increased computational cost. |
| **MLA-EDTCNet + CF** [6] | 2025 | Multi-Layer Attention Encoder-Decoder Temporal CNN for sentiment analysis, combined with **collaborative filtering**. Uses an attentive CNN to classify review sentiment, addresses class imbalance with GAN-based oversampling, then injects sentiment polarity into a CF model for final recommendations. | Outperformed state-of-the-art baselines on Amazon product data in accuracy and F1. The sentiment-enhanced hybrid approach yielded higher user satisfaction by filtering recommendations based on predicted sentiment polarity. |

*Table 1 lists a selection of notable approaches, illustrating the diversity of techniques: from graph neural networks and transformers to hybrid and conversational models. Next, we delve deeper into each category of methods, discussing how they work and what contributions they make.*

## 2.1 Deep Learning Models Integrating Sentiment

The first wave of sentiment-aware recommenders applied **deep learning** to incorporate review information into traditional recommendation frameworks. These models typically extend classic collaborative filtering or factorization techniques by adding a component that learns from review text. Early examples (pre-2020) included CNN or RNN networks that produced latent representations of reviews, which were then fused with latent user/item factors. The core idea is to use an auxiliary neural network (often called a *review encoder*) that transforms each user's written opinions and each item's reviews into vector embeddings. These embeddings are then used in prediction alongside or instead of user–item interaction embeddings.

A representative architecture is the **dual-attention deep model** by *Da'u and Salim (2019)*, which introduced a sentiment-aware neural recommender combining an LSTM(Long Short-Term Memory)-based review encoder with an attention mechanism to focus on important words as said in [5]. They demonstrated that attending to sentiment-bearing words (e.g. "good", "bad", "excellent") in reviews helped predict ratings more accurately [5]. Subsequent works refined this idea by using more advanced encoders and multitask learning. For instance, researchers have used **aspect-level sentiment models** to extract specific features (aspects) from text and align them with user preferences. One model integrated a semi-supervised topic model with an LSTM and co-attention network to jointly learn aspects and sentiment weights [5]. The result was improved prediction performance by aligning users with items on the aspects they explicitly praised or criticized.

By 2021–2022, deep learning models began to leverage **attention mechanisms and memory networks** to better capture long-range dependencies in reviews. Some models treat each review as a sequence and use hierarchical attention: first aggregating words into a review sentiment vector, then aggregating multiple reviews of an item or by a user [5]. This hierarchy matches the data structure (multiple words per review, multiple reviews per entity). Models like NARRE (2018) and its successors learned to attend to the most informative reviews for each user and item, effectively selecting the critical opinions that influence recommendation. These architectures were successful in improving rating predictions and even enabling **explainable recommendations** (by highlighting which review content was pivotal in the model's decision).

A key challenge for deep models is to jointly learn from *two modalities*: structured interaction data (ratings, clicks) and unstructured text (reviews). One approach is to train the networks in a multi-task fashion, predicting both star ratings and review sentiment simultaneously. Another approach is to incorporate the text-derived features directly into the preference prediction model. *Zhao et al. (2020)*, for example, proposed a cross-domain recommender that leveraged a sentiment lexicon and an attention network (CATN) to transfer sentiment knowledge between domains. They found that sentiment information can help mitigate cold-start problems in a new domain by providing clues about user preferences on item aspects [1].

By the early 2020s, these deep learning approaches established that **sentiment analysis can significantly enhance recommendation quality**. However, they often required careful feature engineering (e.g. aspect extraction or lexicon construction) and had limited ability to capture the full context of language, especially nuances like sarcasm or multi-faceted opinions. This set the stage for the next trend: leveraging pre-trained transformer models to better understand and utilize textual data.

## 2.2 Transformer-Based and Large Language Model Methods

Transformer-based models have revolutionized NLP with their ability to capture context and semantics at scale. In the recommendation domain, researchers quickly recognized the value of transformers like **BERT** and **GPT** for encoding review text and even for generating recommendation outputs. Work in 2023–2024 increasingly centers on integrating these powerful language models into sentiment-aware recommenders.

One straightforward application is using pre-trained transformers (such as BERT) as **feature extractors** for review sentiment. For example, Darraz *et al.* (2024) use a BERT model fine-tuned for sentiment classification on Yelp reviews to create sentiment features for users and items [3]. In their framework, each review is fed into BERT to obtain a sentiment-informed embedding (or an explicit sentiment score like positive/negative), and these embeddings are aggregated per user/item. The aggregated sentiment vectors then augment a collaborative filtering model. This method showed clear gains: by focusing on two domains (restaurants and hotels) and leveraging BERT's language understanding, the hybrid model achieved better recommendation accuracy than a baseline without sentiment analysis [3]. The success is attributed to BERT's ability to understand context (e.g. distinguishing *"not great"* from *"great"* in a sentence) and thus provide more reliable sentiment signals than earlier bag-of-words or simple RNN approaches.

Beyond feature extraction, transformers have been used as the **core** of recommendation models. The work by Gheewala *et al.* (2024) conducts a *"compendious survey"* and empirical evaluation of deep learning vs. transformer models for review-based recommendation. Their experiments, which included transformers like RoBERTa, showed that deep transformer models yield more **interpretable and relevant** user/item representations than traditional CNN or RNN networks. In fact, the best transformer model in their study achieved up to ~4.6% improvement in RMSE and ~7.4% in MAE on Amazon Electronics recommendations compared to the best non-transformer deep network as shown in [4]. This highlights how transformers, with their self-attention mechanism, can capture subtle signals (like comparative opinions or context-dependent sentiment) that other models miss.

Another development is incorporating **Large Language Models (LLMs)** such as GPT-3 or ChatGPT into the recommendation process. Gao *et al.* (2023) proposed *Chat-Rec*, a framework where an LLM is prompted with a user's history or a few example reviews to generate recommendations or explanations as displayed in [1]. The LLM can perform sentiment analysis implicitly – for instance, by interpreting a user's review text and then suggesting a suitable item while explaining the reasoning. Large models bring a few advantages: they come with world knowledge and conversational ability, enabling more interactive and context-aware recommendations [10]. They can, for example, understand a complex user query like *"I loved the camera on my last phone but hated the battery life, what do you suggest?"* and parse out the positive and negative sentiments to recommend a phone with a good camera and battery. However, using LLMs also poses challenges such as high computational cost and the need to guard against the generation of irrelevant or biased content.

A compromise solution seen in literature is to use transformers for **representation learning**, then plug those representations into lightweight models for the actual recommendation. For instance, a system might encode all item descriptions and user reviews using a transformer into embeddings offline, and then use those embeddings in a standard nearest-neighbor or matrix factorization approach. This leverages the transformer's strength in NLP while keeping the recommendation stage efficient. *Xiang et al.* (2024) survey exactly this kind of pipeline, reviewing how transformer-based text understanding (for product descriptions, user comments, etc.) can feed into recommendation generation from [9]. They emphasize that pre-trained models help in "understanding complex user intentions and improving the quality of recommendations" [9], especially in cold-start scenarios where only textual data (a new item's description or initial reviews) is available.

In summary, transformer-based methods are becoming a **backbone** for sentiment-aware recommenders. They reduce the need for manual feature engineering by excelling at **contextual sentiment analysis**. They also open up new capabilities like **zero-shot reasoning** (an LLM can make a recommendation in a new domain by analogy to what it knows about language). The trend is moving toward combining the predictive power of traditional recommender algorithms with the nuanced understanding of language from transformers. However, integrating these at scale in real-world e-commerce (with millions of items and users) remains an active area of research, touching on system optimization and efficiency – we will revisit this in the Challenges section.

## 2.3 Conversational and Interactive Systems

Conversational Recommender Systems (CRS) engage the user in a dialogue to elicit preferences and feedback. In a conversational setting, the system may ask questions, make suggestions, and adapt based on the user's responses. **Sentiment analysis plays a crucial role** here, because the user's utterances in the conversation often carry sentiment (e.g. *"I didn't like that recommendation"* is a strong negative signal). By being sentiment-aware, a CRS can adjust its strategy – for instance, apologize and offer a different type of item if the user is unhappy, or focus on what the user expressed liking.

One recent advancement is the *Sentimentally Enhanced Conversation Recommender (SECR)* by Liu *et al.* (2025). SECR addresses shortcomings of prior conversational systems which *"fail to thoroughly explore users' emotional tendencies toward entities"*. In a movie recommendation dialogue, for example, earlier systems might treat a user mentioning a movie as a sign of interest, even if they said they disliked it. SECR fixes this by analyzing the sentiment of each user utterance about movie entities and then **filtering or re-weighting those entities** in the recommendation algorithm. It introduces an **emotional weighting mechanism**: entities (like movie titles, actors, genres) the user speaks about are given an importance score based on the user's sentiment (positive sentiment boosts the entity's score, negative sentiment reduces it). These scores influence which movies the system will recommend next [8]

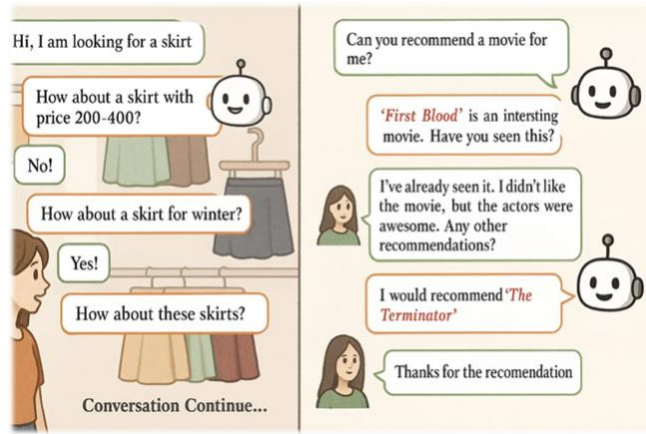

*Figure 2: Example of a conversational recommendation scenario (simplified from Liu et al. 2025). The left shows an attribute-based dialogue for a shopping query, and the right shows a generation-based movie recommendation dialogue. Notably, on the right, the user explicitly says "I didn't like the movie" (negative sentiment about First Blood), but a sentiment-unaware system still recommends a similar movie (The Terminator), leading to a suboptimal suggestion . A sentiment-aware system would recognize the negative feedback and avoid recommending a similar entity.*

SECR's architecture is multi-component. It builds a comprehensive movie knowledge graph (MAKG) to expand the space of entities it knows (movies, actors, etc.), since relying on a fixed knowledge source like DBpedia was found to be incomplete. During a conversation, the system uses **NLP techniques (prompts and lexicons)** to extract sentiment. They specifically use the NRC VAD lexicon and prompt-based sentiment queries to get an "entity sentiment score" for any entity the user mentions. Entities with sentiment below a threshold (e.g. clearly negative opinions) are filtered out from recommendations. The system also brings in external information like brief plot introductions (via a Pegasus summarization model) to enrich the conversation, aiming to keep dialogue responses diverse and engaging. As a result, SECR can avoid mistakes like recommending *The Terminator* after the user said they **disliked** *First Blood* (a scenario explicitly illustrated in their paper). Instead, it might focus on entities the user did show positive sentiment towards (e.g. the user liked the *actors* in *First Blood*, so SECR might recommend another movie with those actors, addressing the nuanced preference) as exhibited in [8].

Beyond academic research prototypes, the idea of **conversational sentiment adaptation** is highly practical. E-commerce chatbots can monitor user emotions (frustration, satisfaction) through sentiment analysis of messages and adjust their tone or strategy. For example, if a user sounds annoyed, the system might switch to a different recommendation category or offer a small incentive. Some recent works (e.g. by Zhang *et al.* 2023 in the dialogue domain) also explore using emotion detection to decide how a conversational agent should respond – possibly recommending a comforting product if user sentiment is very negative, though this area is nascent.

Another aspect is **explanations in conversational form**. Large Language Models have been used to make conversational recommenders not just interactive but also explainable. *Chen et al.* (2023) integrated an LLM to generate explanations that include features and sentiments ("I suggest this laptop because many reviews praise its battery life, which you said is important to you") [15]. The LLM evaluates which product features have positive sentiments from the user or general population and frames the recommendation accordingly. This can increase user trust in the system's suggestions.

In summary, conversational systems benefit from sentiment awareness by becoming more **responsive and user-centric**. They represent a move from one-shot recommendations to a **feedback loop**, where each user utterance (and its sentiment) informs the next system action. However, building such systems requires solving challenges in natural language understanding, maintaining context over multiple turns, and balancing recommendation quality with dialogue flow. The early results from systems like SECR are promising – they show higher user satisfaction and more efficient dialogues (fewer turns needed to reach a good recommendation) when sentiment is taken into account [8].

7## 3     Results – Data & Metrics

Research on sentiment-aware recommenders leverages a variety of datasets that contain both user-item interactions and textual reviews or comments. Table 2 provides an overview of some widely used datasets in this field, highlighting their domain (type of products or services), scale, and relevance.

*Table 2.* Notable Datasets for Sentiment-Aware Recommendation Systems

| Dataset Name & Source | Domain | Scale (Reviews, Users, Items) | Description & Notes |
|---|---|---|---|
| **Amazon Reviews 2023** [7] | E-commerce (All categories) | 571.5M reviews; 54.5M users; 48.2M items | Massive multi-domain dataset (electronics, books, clothing, etc.) with ratings and text. Collected 1996–2023. Rich metadata (price, images) also available. Often used for deep learning benchmarks; poses challenges in scalability. |
| **Yelp Reviews** [14] | Local businesses (Restaurants, etc.) | ~7.0M reviews; ~1.99M users; 150k businesses | Public Yelp dataset (annual release for academic use). Includes star ratings, review text, business attributes, and user info. Reviews often contain sentiment-rich descriptions of service and ambiance. Frequently used for experiments in sentiment analysis and recommendation due to its balanced size and text availability. |
| **TripAdvisor Hotel** [11] | Travel (Hotels) | 0.88M reviews; ~260k users; 4,333 hotels | Hotel review dataset crawled from TripAdvisor. Each review has a 1–5 rating and text; some are multi-sentence detailed opinions (sometimes multilingual). Useful for studying aspect-based sentiment (e.g., cleanliness, location) in recommendations. Smaller in scale, but domain-specific. |
| **MovieLens + Reviews** [19] | Movies/Films | ~1.1M ratings; ~27k users; ~9k movies | MovieLens datasets contain explicit ratings but no text. Researchers often augment them with review text from sources like IMDb. This combination allows testing sentiment-aware models in movie recommendation. The data size is moderate; good for rapid experimentation, but requires aligning external reviews to MovieLens IDs. |
| **Alibaba iFashion** [12] | E-commerce (Fashion outfits) | 0.3M outfits; item attributes; some user interactions | A specialized dataset with product images, descriptions, and some user behavior (from Alibaba). While focused on outfit compatibility, it has user comments with sentiments on style. Illustrates use of multimodal sentiment (text + image) in recommendations, though not as common as pure text datasets. |

*Table 2* lists datasets spanning large general-purpose collections (Amazon) to domain-specific ones (hotels, fashion). The **Amazon Reviews** dataset, originally compiled by Julian McAuley's group, is perhaps the most influential – its sheer size and diversity have driven development of scalable models. It enables cross-domain evaluation (e.g., training a model on Electronics reviews and testing on Books) and analysis of cold-start scenarios for new products using text. However, working with hundreds of millions of reviews is computationally intensive, so researchers often use subsets (for example, focusing only on specific categories like Electronics or Beauty). Evaluation on Amazon data typically uses metrics like RMSE for rating prediction or Hit Rate for top-N recommendation, comparing methods with and without sentiment features.

8The **Yelp** dataset is another benchmark – its reviews are longer and more opinionated (on average) than Amazon's, since people often narrate their experience. It also provides context like geographic location and social connections (friends), enabling studies that combine sentiment with social filtering (as in some graph-based approaches). Many sentiment-aware recommender papers report results on Yelp for tasks like rating prediction or business ranking. A challenge with Yelp is that it has a lot of **unstructured information** (e.g. check-in data, photos) that some models ignore; focusing only on text and ratings provides a clean comparison ground for NLP-infused recommenders.

The **TripAdvisor** and **MovieLens+IMDb** examples represent smaller datasets useful for qualitative analysis and cold-start tests. For instance, a model can be trained on TripAdvisor reviews to see if it can recommend hotels based on certain aspects (like "quiet room" or "friendly staff" mentions). Smaller datasets also allow for human evaluation of recommendations – e.g. verifying if a sentiment-aware model indeed picks up relevant phrases from reviews when making recommendations.

In evaluation, researchers commonly use **Rating Prediction** and **Top-N Recommendation** as the two main tasks:

- *Rating Prediction:* Predict the numerical rating a user would give to an item. This is where RMSE, MAE are used. Sentiment-aware models aim to lower these errors by using review text to predict ratings closer to ground truth.

- *Top-N Recommendation:* Provide a ranked list of N items a user is most likely to interact with positively. Here metrics include Precision@N, Recall@N, NDCG, or Hit Rate. Incorporating sentiment can change the ranking significantly, ideally improving the hit rate of items that the user actually likes (as evidenced by their past review sentiments).

*Standardized Evaluation and Reproducibility*

Despite the varying scales and domains—from the massive, multi-category Amazon Reviews to the more focused TripAdvisor hotel and Alibaba iFashion collections—researchers increasingly recognize the need for standardized, reproducible evaluation pipelines. Toolkits such as RecBole have emerged to consolidate review-based models and benchmarking scripts, helping ensure that improvements attributed to sentiment integration are not confounded by inconsistent data splits or preprocessing choices. Unified benchmarks facilitate fair comparisons across methods—from simple bag-of-words to advanced transformer-based and graph-augmented architectures—and, as models grow in complexity, reproducibility and standardized evaluation become critical; surveys have highlighted the need for consistent baselines, publicly released code and models, and unified benchmarks [13]. While RMSE and Hit Rate remain valuable for large-scale datasets like Amazon and Yelp, they can mask limitations in recommendation diversity and novelty; to address this, some studies complement quantitative evaluation with qualitative user studies and case analyses that assess explanation coherence *(e.g., "Recommended because you mentioned you love battery life in phones")* using BLEU scores or human judgments, and enable qualitative assessments—such as coherence of generated explanations—across smaller, domain-specific collections.

With the emergence of large language models (LLMs) capable of producing nuanced, user-oriented explanations—such as "Recommended because you prefer long battery life"—transparency in experimental protocols becomes essential. Researchers must disclose exact model checkpoints, prompt templates, decoding parameters (e.g., temperature, top-k/top-p), and fine-tuning procedures. Experiment-tracking tools like MLflow, Weights & Biases should log random seeds, hyperparameters, data splits, and environment configurations. Containerization with Docker or Singularity and dependency manifests (requirements.txt, conda.yml) guarantee reproducibility. Equally vital is the release of consistent, shared baselines and unified benchmarks: training and evaluation scripts should reside in public repositories under permissive licenses, accompanied by preprocessed data splits or recreation instructions and pretrained model weights. Leaderboard-style benchmark suites—hosted on platforms like Hugging Face datasets and evaluation hubs—enable statistical significance testing and transparent comparison across methods. These practices foster a collaborative culture in which innovations in sentiment-infused recommendation and explanation generation can be validated, compared, and built upon.

# 4 RESULTS – CHALLENGES & INSIGHTS

Despite the notable advancements, there remain several **open challenges** in developing sentiment-aware recommendation systems:

4.1 **Noise and Ambiguity in Text:** User-generated content can be messy. Slang, sarcasm, and typos can mislead sentiment analysis. For instance, a review stating *"I wouldn't recommend this laptop to **anyone**"* is clearly negative, but a bag-of-words model might misinterpret "recommend" as positive. Irony and subtlety are still difficult for models, even transformers, to fully grasp in every case. Misinterpreting sentiment can be worse than ignoring it – a recommender might wrongly assume a user liked something they actually hated. Developing robust sentiment analysis (possibly with sarcasm detectors or by leveraging the context of multiple user reviews) is an ongoing challenge.

4.2 **Aspect-Level Sentiment Alignment:** Users often express opinions by referencing item features, but aligning these aspect-level sentiments with recommendation decisions remains challenging. For example, balancing sound quality and comfort preferences in suggesting another pair of headphones is difficult. Current models reduce sentiment to a single scalar or manage only a few aspects. The challenge is transitioning to more detailed representations, where user preferences are expressed as vectors of sentiments across various aspects. Attention mechanisms and aspect-based sentiment analysis have shown progress, but a comprehensive approach that integrates aspect sentiment into the core recommender algorithm is still developing.

4.3 **Dynamic and Contextual Preferences:** Sentiments and opinions can change over time or with context. Traditional recommenders struggle with these shifts. Sentiment-aware systems could adjust if they notice positive reviews for a disliked genre. However, temporal modeling is required. Zhang et al. (2023) introduced a dynamic graph for sentiment, but few recommender systems fully utilize it. Incorporating time-aware sentiment analysis to handle concept drift and seasonal trends is needed.

4.4 **Cold Start for New Items/Users:** Cold start remains challenging. Sentiment analysis on available text (e.g., product descriptions, early reviews) can signal sentiment for new items. However, models need confidence in these signals. External data (social media, product descriptions, competitor reviews) can infer sentiment for cold-start items. Leveraging user reviews on other platforms or sentiment profiles can also help. Cross-domain sentiment transfer is an area for growth, e.g., using Yelp review sentiments to warm-start Amazon product recommendations.

4.5 **Scalability and Efficiency:** Incorporating text, especially with deep models like transformers, is computationally expensive. Techniques like distillation, approximate nearest neighbor search, and offline recommendation updates help, but every added complexity must justify its cost. Optimizing systems involves indexing sentiment-enriched representations and updating models with new data without re-training. Researchers explore incremental learning and lighter architectures, but more work is needed to balance response times and memory usage with production requirements.

4.6 **Evaluation Metrics and Benchmarks:** As hinted earlier, a gap exists in how we evaluate success. Most works focus on accuracy metrics (RMSE, Precision@N), but improvements there don't always translate to better user satisfaction. We lack standardized ways to measure the **quality of explanations** or the **user experience** with a sentiment-aware system. Also, the community would benefit from more public benchmarks that include both rating and text, with clear tasks defined (some progress here with datasets like Amazon and Yelp, but perhaps a unified challenge or leaderboard could drive further progress). Additionally, incorporating A/B testing results or user study results in research papers could bridge the gap between offline metric improvements and real-world impact.

4.7 **Bias and Fairness:** User reviews and sentiments can reflect societal biases, such as unjust harsh reviews for businesses in certain neighborhoods or gendered language affecting sentiment interpretation. Sentiment-aware recommenders may inadvertently amplify biases, like under-recommending minority-owned businesses due to skewed sentiment. Researchers should consider fairness-aware training and debiasing sentiment analysis to address these issues. Privacy implications arise from using personal review data, and future systems may need privacy-preserving sentiment analysis methods like on-device processing or differential privacy.

The field lacks research on improving sentiment integration's fineness, adaptability, and reliability. Current models excel in labs but need refinement for real-world e-commerce. Addressing these challenges is crucial for next-gen recommender systems to truly understand and cater to users' preferences.

# 5   Discussion

Building on the challenges identified above, we propose the following research directions:

5.1 **Multi-Modal Sentiment-Aware Recommendation**: Although current systems largely depend on textual sentiment, user opinions are also expressed through modalities such as images, videos, or voice. Future models can enhance recommendation quality by combining text-based sentiment analysis with visual sentiment recognition, for example, by interpreting emotions from user-posted product images or videos. Datasets such as Amazon and Yelp, which offer both text and visual content, present an opportunity to investigate whether integrating image features with textual reviews results in more personalized recommendations. This direction calls for the convergence of NLP and computer vision to extract richer sentiment signals.

5.2 **Cross-Lingual and Multilingual Sentiment Integration:** E-commerce operates on a global scale, making it essential to understand sentiments expressed in multiple languages. However, most existing research is limited to English-language datasets. Future work could extend sentiment-aware recommendation by incorporating multilingual or translated review data. This might involve using translation APIs or pre-trained multilingual models like multilingual BERT to create a more inclusive system. A model that leverages English reviews to help interpret sentiment in another language would broaden applicability to diverse markets.

5.3 **Personalized Sentiment Analysis:** Sentiment expression varies significantly across users. Some individuals use restrained language when expressing dissatisfaction, while others may exaggerate praise or criticism. A generalized model may fail to capture these differences. Personalized sentiment analysis offers a way to tailor sentiment models to individual linguistic styles. This could involve clustering users based on their review behavior or embedding user-specific traits into sentiment classifiers. Incorporating user embeddings could help the system learn how a specific user expresses sentiment, enhancing predictive accuracy.

5.4 **Causal Inference for Sentiment Influence:** Sentiment-aware models demonstrate promising performance, but their effectiveness hinges on correlation rather than causation. Future research endeavors should explore the direct impact of sentiment features on recommendation outcomes. For instance, does a mention of "excellent battery life" directly influence the model's recommendation? Causal inference methodologies and experimental approaches can facilitate the elucidation of these intricate relationships.

5.5 **Human-in-the-Loop and Interactive Tuning:** User-centric design is critical for building adaptable recommendation systems. Future research could focus on enabling users to adjust sentiment-driven filters dynamically. Consider an interface where users express preferences such as "prefer items with high cleanliness scores" or "avoid low-rated products." This can be achieved through real-time tuning with interactive controls, such as sliders. For instance, a travel recommendation system might let users emphasize different aspects—like location or service—based on past review sentiments. This approach integrates user experience design with algorithmic responsiveness, promoting more interpretable systems.

5.6 **Integration with Large Language Models for Explanations:** Large language models (LLMs) can generate coherent, sentiment-grounded explanations for recommendations. A promising direction involves fine-tuning LLMs to generate explanations based on user reviews. By training models on datasets containing user-item-review triplets mapped to rationales, LLMs can articulate recommendation logic in user-friendly terms. For example, an LLM might explain a product suggestion by referencing prior positive sentiments on similar features. Retrieval-augmented generation can further ground these explanations in actual user reviews, boosting user trust.

5.7 **Benchmarking and Reproducibility Initiatives:** Despite growing interest in sentiment-aware recommenders, the field lacks standardized evaluation benchmarks. A critical step is establishing shared tasks, datasets, and evaluation protocols. Organizing challenges around problems such as cold-start scenarios or cross-lingual sentiment understanding can foster more consistent evaluation. Even without new algorithms, efforts in dataset curation, evaluation code development, and baseline comparisons can significantly enhance reproducibility. These contributions enable fair comparisons and encourage adoption in real-world applications.

Collectively, these directions reflect a shift toward building more flexible, responsive, and user-aware recommendation systems. They encourage cross-disciplinary research, drawing from NLP, computer vision, causal reasoning, and interactive systems. By exploring these areas, future work can help design recommendation engines that are not only accurate but also interpretable, inclusive, and aligned with diverse user preferences.

## 6  CONCLUSION

In this review, we surveyed the evolving landscape of **sentiment-aware recommendation systems** in e-commerce, focusing on developments from an NLP perspective in the past few years. We began by motivating why infusing recommendations with sentiment analysis of user feedback is valuable for personalization and user satisfaction. The integration of NLP techniques enables recommender systems to go beyond star ratings and tap into the *reasons* behind user preferences, making suggestions that align more closely with a user's expressed opinions.

We then discussed a range of approaches – from deep learning models that encode review text alongside ratings, to cutting-edge transformer-based architectures that leverage pre-trained language models, to graph neural networks that incorporate sentiment into user-item relationships, and finally to conversational systems that dynamically adjust to user sentiment in dialogue. Through illustrative examples and figures, we saw concrete scenarios (like **Figure 2's** conversation) where a sentiment-aware approach clearly outperforms a sentiment-blind one by avoiding pitfalls of misunderstanding user feedback. We also summarized representative models (Table 1) and datasets (Table 2) to provide a quick reference of the state-of-the-art methods and the resources available for evaluation.

Our review highlights that **major trends** in 2023–2025 include the use of transformer models for deeper text understanding and the blending of multiple sources of data (text, graphs, interactions) to capture complex user preference signals. There is a visible shift towards making recommendations more transparent and explainable using the content of user reviews. We identified several **research gaps** such as handling noisy language, aligning aspect-level sentiments with recommendations, adapting to temporal changes, and ensuring scalability. These challenges temper the successes achieved so far and point to where further research is needed. The **future directions** we suggested aim to guide upcoming work to address these gaps and explore new frontiers – including multi-modal sentiment analysis, personalization of sentiment interpretation, and interactive user controls for preference tuning. By following these directions, researchers can build on the foundation laid by recent advances and create next-generation recommenders that are emotionally intelligent, context-aware, and user-centric.

In conclusion, sentiment-aware recommendation is a vibrant interdisciplinary area bridging NLP and recommender systems. It holds great promise for making e-commerce platforms more **responsive to user needs** – not just recommending highly-rated products, but the *right* products for each user, backed by an understanding of that user's unique likes and dislikes. As the volume of online reviews and opinions continues to grow, these systems will be crucial in distilling opinion data into actionable personalization. We expect that sentiment-aware recommenders will increasingly move from research prototypes to real-world deployments, improving the quality of our interactions with digital commerce. The progress reviewed in this paper demonstrates that when recommendation algorithms truly "listen" to what users say (and feel), the outcomes are beneficial for users and platforms alike – delivering recommendations that resonate on a deeper level than ever before.

## 7  DECLARATIONS

**Author Contributions:** Yogesh Gajula (Y.G.) conceived the review topic, conducted the literature search covering publications from 2023 to 2025, evaluated and synthesized the key findings, and drafted the manuscript.

**Conflict of Interest:** The author declares no competing interests.

**Ethics Approval and Consent to Participate:** This work is a literature review and did not involve human participants, animals, or personal data; no ethical approval or consent was required.

**Funding:** This research received no external funding.